\documentclass[universe,article,accept,pdftex,moreauthors]{Definitions/mdpi} 

\firstpage{1} 
\pubvolume{9}
\issuenum{8}
\articlenumber{372}
\pubyear{2023}
\copyrightyear{2023}
\externaleditor{Academic Editor: Firstname Lastname}
\datereceived{5 July 2023} 
\daterevised{27 July 2023} 
\dateaccepted{ } 
\datepublished{ } 
\hreflink{https://doi.org/10.3390/universe9080372} 

\Title{Galaxy Rotation Curve Fitting Using Machine Learning Tools}

\TitleCitation{Rotation Curve Fitting Using Machine Learning Tools}

\Author{{Carlos. R. Arg\"uelles} $^{1,2,}$*$^{,\dagger}$\orcidA{} and {Santiago Collazo} $^{1,}$*$^{,\dagger}$} 

\AuthorNames{C. R. Argüelles, S. Collazo}

\AuthorCitation{Argüelles, C.R.; Collazo S.}

\address{%
$^{1}$ \quad Instituto de Astrof\'isica de La Plata, UNLP-CONICET, Paseo del Bosque s/n, La Plata B1900FWA, Argentina\\
$^{2}$ \quad ICRANet, Piazza della Repubblica 10, I-65122 Pescara, Italy
}

\corres{Correspondence: carguelles@fcaglp.unlp.edu.ar (C.R.A.); scollazo@fcaglp.unlp.edu.ar (S.C.)}

\firstnote{{These authors contributed equally to this work.} 
}

\abstract{Galaxy rotation curve (RC) fitting is an important technique which allows the placement of constraints on different kinds of dark matter (DM) halo models. 
In the case of non-phenomenological DM profiles with no analytic expressions, the art of finding RC best-fits including the full baryonic $+$ DM free parameters can be difficult and time-consuming. In the present work, we use a gradient descent method used in the backpropagation process of training a neural network, to fit the so-called Grand Rotation Curve of the Milky Way (MW) ranging from $\sim$1 pc all the way to $\sim$$10^5$ pc. We model the mass distribution of our Galaxy including a bulge (inner $+$ main), a disk, and a fermionic dark matter (DM) halo known as the Ruffini-Arg\"uelles-Rueda (RAR) model. This is a semi-analytical model built from first-principle physics such as (quantum) statistical mechanics and thermodynamics, whose more general density profile has a \textit{dense core} -- \textit{diluted halo} 
 morphology with no analytic expression. As shown recently and further verified here, the dark and compact fermion-core can work as an alternative to the central black hole in SgrA* when including data at milliparsec scales from the S-cluster stars. Thus, we show the ability of this state-of-the-art machine learning tool in providing the best-fit parameters to the overall MW RC in the $10^{-2}$--$10^5$ pc range, in a few hours of CPU time.}

\begin{document}


\section{Introduction}\label{sec:1}


Disk galaxies, like our own, are rotational supported structures with the advantage of having baryonic (or luminous) mass tracers in approximate circular orbits from which it is possible to obtain the so-called RC. The specific DM distribution, usually dubbed as the DM density profile, is inferred by fitting the observed velocity RC as a function of the galactocentric radius. Typically, this is carried out by assuming a given underlying DM profile together with different mass models for the luminous components such as the bulge, disc, etc. (see, e.g., \cite{2017PASJ...69R...1S} for a review). Most of these studies assume phenomenological DM profiles (in spherical symmetry) obtained from classical N-body cosmological simulations with a given analytic expression, besides the visible mass components. However, other kinds of DM profiles can be obtained from first-principle physics (i.e., thermodynamics and statistical mechanics) while accounting for the quantum nature of the particles, such as the RAR model for fermions 
(see \cite{2023Univ....9..197A} for a review, {and references therein)
} and, e.g., \cite{2018MNRAS.475.1447B,2019MNRAS.483..289R} for bosonic DM, with no analytic expressions for the profiles. 

The RAR model consist in a self-gravitating system of fermions in General Relativity, and therefore is built upon a coupled system of (ordinary) highly non-linear differential equations, which defines a boundary condition problem to be solved numerically (see, e.g., \cite{2015MNRAS.451..622R} for its original version and \cite{2018PDU....21...82A} for its more realistic extension including for particle evaporation). In its extended version, the RAR model involves four free parameters: $m$ the particle mass, and the set ($\beta$, $\theta$, $W$) of dimensionless parameters reading for the temperature, degeneracy, and cut-off particle energy, respectively. These parameters are present in the underlying coarse-grained distribution function (DF) of the particles (which is of Fermi--Dirac type as explained in Section \ref{sec:2b}), and have to be set at the center of the configuration (denoted with the subscript $0$) in order to solve the system of equilibrium differential equations of the RAR model (see Equations (9)--(13) in \citet{2018PDU....21...82A}). 

When applied to real galaxies, in the recent past the RAR equations were solved for given boundary halo conditions taken from observations. For example, when applied to the MW in \cite{2018PDU....21...82A}, three boundary conditions were considered for the overall DM halo mass: one at the fermion-core ($M_c=4.2\times 10^6 M_\odot$, in order to be an alternative to the BH in SgrA*), and the other two at mid-outer DM halo (i.e., $M(r=12 \rm kpc)\approx 5\times10^{10} M_\odot$ \cite{2013PASJ...65..118S}, and $M(r=40 \rm kpc)\approx 2\times10^{11} M_\odot$ \cite{2014MNRAS.445.3788G}, respectively). Since the RAR model has four free-parameters, once the particle was fixed within the range\endnote{{The} 
 compactness of the fermion-core is inversely proportional to $m$ \cite{2018PDU....21...82A}, and thus it is shown that for $m<48$ keV the core is too extended to fit within the S-2 star pericenter, while for $m>345$ keV the solutions are unstable since the critical value for collapse to a BH is reached at $m=345$ keV.} $(48,345)$ keV, there exists one \textit{core-halo} solution for such a particle mass (i.e., three boundary conditions for three remaining free-parameters) fulfilling with the constraints. Interestingly, even if only three boundary conditions were used from observations at very different scales, the overall behavior of such a RAR DM solution is good enough to fit within the error bars of the Grand RC \cite{2018PDU....21...82A} (after standard baryonic mass models are included). 

A more refined phenomenological analysis of the relativistic RAR model would require a best-fit procedure using the full data points of the corresponding RC including their errors (e.g., using MCMC or grid-coverage methods). Such kind of analysis has recently been performed in \cite{2023ApJ...945....1K} within an MCMC method for a large sample of $120$ galaxies of the SPARC catalog \cite{2016AJ....152..157L}, with explicit $\chi^2$ minimization and corresponding posteriors for the RAR model parameters for a fixed particle mass fixed at $m=50$ keV. However, besides the fixed $m$ case analyzed in \cite{2023ApJ...945....1K}, the baryonic mass models were fixed according to the SPARC-catalog (for each galaxy). Thus, it is of interest to develop a numerical technique which, for non-analytic DM models such as RAR, makes it possible to provide best-fits to the full RC-data when including for a larger free parameter-space (i.e., full DM $+$ baryonic model parameters) in a few hours of CPU-time. 

Thus, in this work we propose a new RC best-fitting method based on state-of-the-art machine learning tools, when including for baryonic free parameters together with the full four free-RAR model parameters. In particular we will focus our attention in the so-called Grand RC as studied in \cite{2017PASJ...69R...1S} further including for with innermost data coming from the S-cluster stars orbiting SgrA*, thus covering in total about seven orders of magnitude in galactocentric-radius (e.g., within $\sim$$10^{-2}$--$10^5$ pc). All in all,  the ability of the RAR model to provide excellent fits to the MW RC covering very different radial-scales is shown, involving large order-of-magnitude variations in the gravitational potential generated by the overall total mass distribution of the Galaxy.

\section{Rotation Curve Data and Methodology}
In this section we describe the data selection together with the methodology. We first detail the different data points considered in this work, both coming from the S-cluster stars around SgrA* as analyzed in \cite{2018PDU....21...82A} from data taken in \cite{2009ApJ...707L.114G}, and the ones coming from Pop. I stars and interstellar gas as compiled in \cite{2017PASJ...69R...1S} including for different observational techniques. Then we make explicit the different mass models here assumed, accounting for a central dark compact object (no-BH), inner $+$ main bulge, disc, and outer DM halo. It is important to emphasize once more that both the massive dark compact object together with the outer halo are different components of the very same fermionic RAR DM-model; that is, the dense fermion-core (supported against gravity by fermion-degeneracy pressure) is surrounded by a more dilute DM halo (supported against gravity by thermal pressure). Finally, we briefly describe the machine learning tool methods applied here to make the best fit for the given Grand RC.

\subsection{Data Selection}
\begin{figure}[H]
    \includegraphics[width=12.cm]{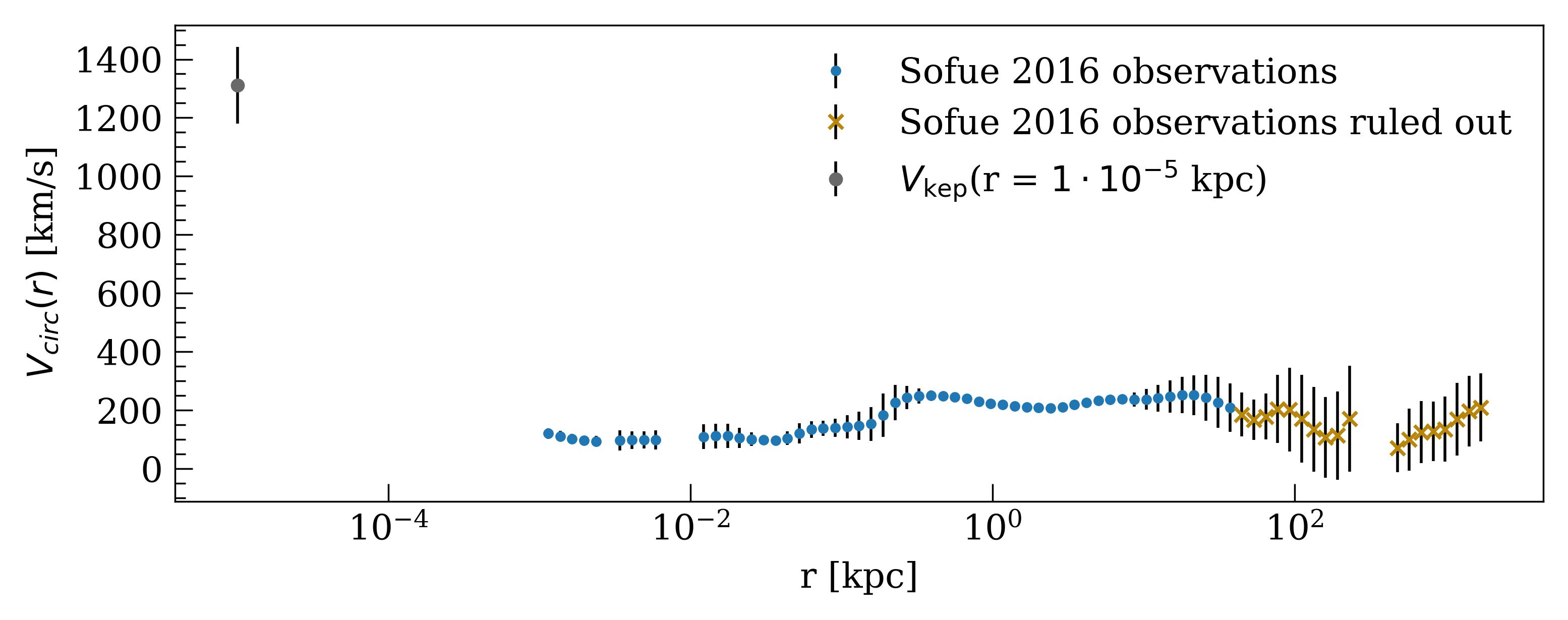}
    \caption{Overall Milky Way RC used to constrain the gravitational potential of the Galaxy. It is composed by the Grand Rotation Curve from \cite{2017PASJ...69R...1S} (light-blue and orange dots) and an inner (milliparsec scale) keplerian velocity (grey dot) of an S-cluster star as caused by the central object in SgrA*.}
    \label{fig2}
\end{figure}

We use the observed Grand RC of the Milky Way as provided in \cite{2013PASJ...65..118S,2017PASJ...69R...1S}, including for different combined observational methods (with associated systematics), ranging from \mbox{$\sim$1 pc} up to $\sim$$10^5$ pc. Due to large error bars arising above $\approx$40 kpc, we will consider here the overall (processed) RC data\endnote{\url{http://www.ioa.s.u-tokyo.ac.jp/~sofue/htdocs/2017paReview/} (accessed on 15 July 2022)} up to that radial scale (see Figure \ref{fig2}). This RC uses, as Galactic constant values, $R_0=8$ kpc and $V_0=238$ km/s (with $R_0$ the distance of the sun from the Galaxy center and $V_0$ the circular velocity of the Local
Standard of Rest (LSR) at the Sun). Additionally, we follow the analysis of the eight best-resolved S-cluster stars \cite{2009ApJ...707L.114G} as used~in~\cite{2018PDU....21...82A} by considering their average circular-orbit velocity. As clearly shown in \cite{2018PDU....21...82A}, they follow the expected Keplerian velocity trend as a function of galactocentric radius due to the gravitational potential of the dense central object, all the way up to $\sim$$10^{-1}$ pc  \cite{2013PASJ...65..118S,2017PASJ...69R...1S}. Thus for convenience we will choose a typical (average) circular-velocity value within such an inner Keplerian trend at $10^{-2}$ pc (see innermost data point in Figure~\ref{fig2}). The combination of such different radial scales covered in the data here selected is, of course, motivated in view of the \textit{core-halo} nature of the RAR DM model, for which a best-fit in their free parameters will be attempted.

\subsection{Mass Models of the MW}\label{sec:2b}
The full dynamical parameters for an accurate Galactic mass determination is very complex. As detailed in \cite{2017PASJ...69R...1S}, they can be divided into three main categories: an axisymmetric structure (or RC); a non-axisymmetric structure (outside the RC) including inner bars and arms; and a radial flow (out of the RC) including ring structures. The work~of~\cite{2017PASJ...69R...1S} centered the attention in the RC only, as  is also the case of the present paper. This axisymmetric structure includes  a dark central object, a bulge (with two components), a flat disc and an spherical DM halo. Thus, we model the gravitational potential of the Milky Way through the following different components:

\textbf{{(i)} 
} The bulge is going to be modeled by two exponential spheroids according to \cite{2013PASJ...65..118S,2017PASJ...69R...1S}, one to model the inner bulge and the other to model the main bulge. The matter density for such models, each one providing two free parameters, is:
\begin{equation}
    \label{eqn:bulge}
	\rho(r)=\rho_c\exp{(-r/a_b)},
\end{equation} 
where $a_{b(i)}=3.5\times10^{-3}$ kpc; $a_{b(m)}=1.2\times10^{-1}$ kpc; $\rho_{c(i)}=3.7\times10^{13} M_\odot/\rm kpc^3$ and $\rho_{c(m)}=2.1\times10^{11} M_\odot/\rm kbulgebulgepc^3$, with the sub-indices $i$ and $m$ indicating the inner and main components. While the bulge parameters will be kept fixed in this work for definitness, the free parameters of the disk will be varied along the RAR model parameters to provide the best fit. 

\textbf{{(ii)}} The disk is going to be modeled by an exponential flat disk as studied in \cite{2013PASJ...65..118S}. The surface mass density of such a disc provides for two free parameters ($\Sigma_d$, $a_d$), and reads

\begin{equation}
    \label{eqn:disk}
	\Sigma(R)=\Sigma_d\exp{(-R/a_d)},
\end{equation} 
where $R$ is the standard cylindrical radius, and ($\Sigma_d$, $a_d$) to be determined by our best fitting~procedure. 

\textbf{{(iii)}} Both the dark central compact object together with the DM halo will be modeled by the semi-analytical (extended) RAR model, which, as explained in the Introduction, is based on a self-gravitating system of fermions at finite temperature including for escape of particles and central fermion-degeneracy. This DM model was extensively studied in \cite{2018PDU....21...82A,2023Univ....9..197A} and references therein for the Milky Way, and in \cite{2023ApJ...945....1K} for other galaxy types. It has four free-parameters $(m,\theta_0, W_0, \beta_0)$ with $m$ the DM, particle mass, and ($\beta_0$, $\theta_0$, $W_0$) the dimensionless parameters evaluated at the origin, reading for the temperature, degeneracy, and cut-off particle energy, respectively. The free-RAR model parameters enter in the underlying phase-space DF of the fermions at (quasi) equilibrium, whose formula is given in Equation~ (\ref{fcDF}) below. Interestingly, it can be demonstrated \cite{1998MNRAS.300..981C,2004PhyA..332...89C} that such a Fermi--Dirac-like DF is a quasi-stationary solution of a kinetic theory equation (of Fermionic--Landau form) via the application of a maximum entropy principle. Thus it is a most-probable coarse-grained DF at violent relaxation, extending the original results of Lynden--Bell on the subject.  

\begin{equation}
\bar{f}(\epsilon\leq\epsilon_c) = \frac{1-e^{(\epsilon-\epsilon_c)/kT}}{e^{(\epsilon-\mu)/kT}+1}, \qquad \bar{f}(\epsilon>\epsilon_c)=0\, ,
\label{fcDF}
\end{equation}
where $\epsilon=\sqrt{c^2 p^2+m^2 c^4}-mc^2$ is the particle kinetic energy, $\mu$ is the chemical potential with the particle rest-energy subtracted off, $T$ is the effective temperature, $k$ is the Boltzmann constant, $c$ is the speed of light, and $m$ is the DM fermion mass. The already defined set of dimensionless-parameters are $\beta=k T/(m c^2)$, $\theta=\mu/(k T)$ and $W=\epsilon_c/(k T)$, respectively. It has been further shown \cite{2021MNRAS.502.4227A} that DM halos built upon such a fermionic DF can be done within a Warm DM cosmological framework, and naturally leading to stable halos which can be extremely long-lived with key implications to the formation and further growth of supermassive BHs in the early Universe \cite{2023Univ....9..197A}.

\begin{figure}[H]
    \includegraphics[width=7.cm]{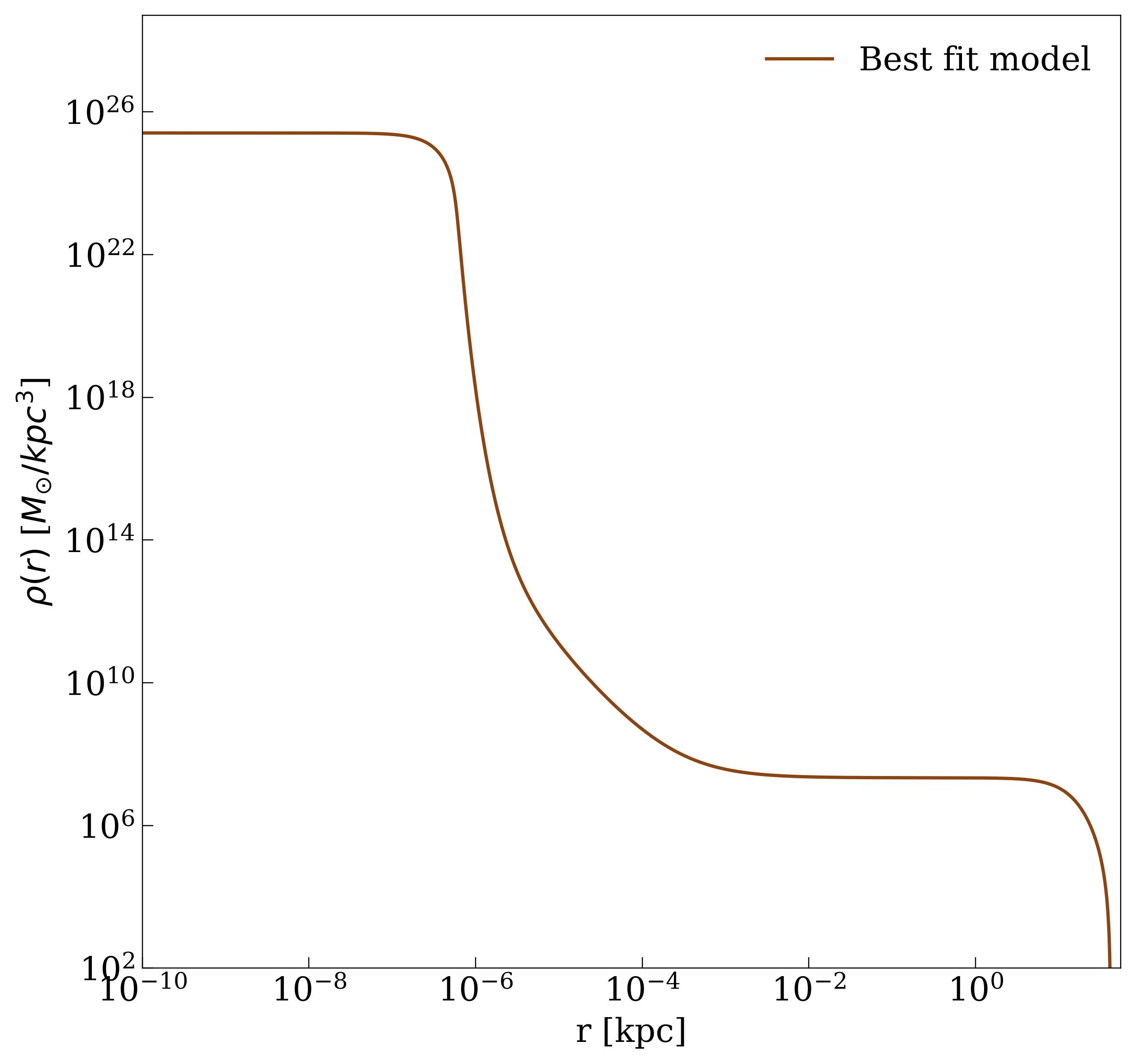}
    \caption{Density profile of the RAR model constrained in this work by the gradient descent method explained in the main text. It can be seen the \textit{constant-density} core below the mili-pc scale (which is governed by quantum degeneracy pressure) and the transition to the plateau at $\sim$1 pc (where quantum effects are negligible), while far above $\sim 1$ kpc, it follows a polytropic tail (see also \cite{2023ApJ...945....1K}).}
    \label{fig1}
\end{figure}

This kind of system, in its most general morphology, develops a density profile with a \textit{dense core}--\textit{diluted halo} distribution (see Figure \ref{fig1}), and is supported against gravity through Fermi degeneracy pressure (for the core) and by thermal pressure (in the outer halo).

\subsection{Gradient Descent Method: A Machine Learning Tool}

\begin{figure}[H]
    \includegraphics[width=11.5cm]{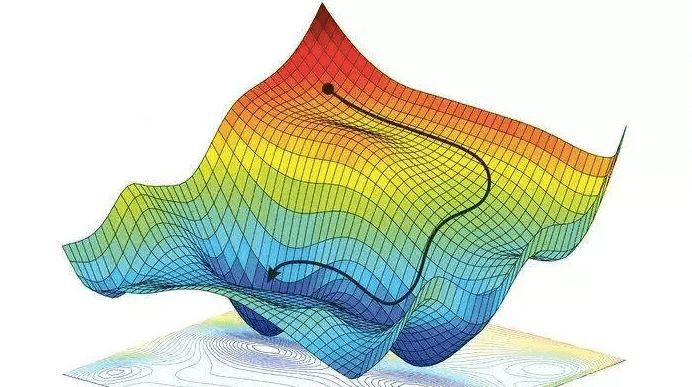}
    \caption{Illustration of the path followed by the gradient descent method to reach the minimum of a two-dimensional function. Taken from: \url{https://easyai.tech/en/ai-definition/gradient-descent/} {(accessed on 10 May 2023).} 
}
    \label{fig:grad_desc_ilus}
\end{figure}

The gradient descent method is based on a progressive sequence of steps to minimize a function. Given a function F to be minimized, it will implement the {formula} 
\begin{equation}
    \mathbf{p^{new}}=\mathbf{p^{old}}-\gamma\nabla_\mathbf{p}F(\mathbf{p^{old}})
    \label{eq:p_new}
\end{equation}
where \textbf{p} stands for the independent variable of the function F and $\gamma$ is a parameter called learning rate whose aim is to regulate the “length” of the steps. If this formula is implemented recursively, one can eventually  go closer and closer to a minimum of F. An illustration of the procedure followed by the gradient descent method is shown in Figure~\ref{fig:grad_desc_ilus}. In that image can be seen a solid dark path, which is the result of evaluating F in the different points given by Equation (\ref{eq:p_new}). Since such a formula uses ``minus'' the gradient of the function, the path followed by the method is oriented to the direction of ``maximum'' decreasing, driving to the deepest point of F.

In the following we provide some pros and cons when using this algorithm. The main advantage is that it finds the minimum of F in a more direct manner (and usually more precisely) than an MCMC or grid-coverage methods: it does not need to ``explore'' a huge volume of different values of F to see which is the minimum one. Instead, it starts to \textit{walk} in a direction and, after iterating the steps, it directly goes towards the minimum. Because of this, it can happen that the algorithm becomes stuck in a local minimum rather than in the \textit{global} one. This is a disadvantage of the method and there are some techniques to avoid this local minima. In our case, we chose to use the gradient descent method as a fine-tuning algorithm of a method providing a ``partial'' minimum of F. That is to say, we took as the initial seed for the gradient descent method the final result of another general optimization method--- that is a genetic algorithm as implemented in the optimized RAR code in PYTHON which will be publicly available through Github in 2023--- 
 the latter making a faster (though less accurate) exploration of the parameter space. 

It is important to emphasize that the idea of this paper is to present a neural-network-prepared tool so it can fit a non-linear non-analytic model with several free parameters like the one here shown. It was not our intention to perform a deep analysis of the fitting technique and/or give a detailed comparison with other tools. So, since there is no ideal way of dealing with the local minima problem, also in neural networks, what we have planned was to apply this method to perform a 'fine-tuning' of the fitting problem once a good-enough initial parameters-seed is provided by other commonly used method (e.g., a genetic algorithm). Thus this fine-tuning technique can be used by anyone who intends to improve the accuracy in the fitted parameters, in highly non-linear models such as, for example, those involving General Relativistic Einstein equations as in the RAR model. With respect to the time savings, the most important comparison is between this method and a traditional MCMC fitting approach. 
While this method took us 1.5 or 2 h to complete the fine-tuning of the free parameters based on a good-fit seed of them, the MCMC approach we have tried before took us several hours or even more (depending on the chains), and only for two free parameters; in this case, we could fit six free~parameters.

Another important comment has to do with a useful advantage related to the morphology of the gradient descent method, since it resides on computing a gradient, and therefore it is less sensitive to the dimension of the parameter space than the MCMC or grid-coverage methods. The latter methods have to walk around all over the subspace of parameters, increasing their computing time drastically for high-dimensional spaces. Indeed, while a very precise RC best-fit for the Milky Way under the same RAR model as used here (with varying baryonic parameters) can take more than a day with genetic algorithms (similar to MCMC), the method  implemented here takes a few hours of CPU-time. 

In the case of this work, the function F introduced above will be a function that quantifies how good  the predictions made by the model are in contrast to the observations. In machine learning, these kinds of functions are called \textit{loss functions}. Specifically, we will use a Mean Squared Error (MSE) as a loss function, which is defined as:
\begin{equation}
    \label{eqn:method:Loss}
	Loss(\mathbf{p}) =1/C \sum \limits_{i=1}^N \frac{\left(V(r_i,\mathbf{p}) - v_i\right)^{2}}{N},
\end{equation} 
where \textbf{p} is the vector of the (physical) free parameters that characterize the full mass model (baryons $+$ DM). In this work we fit six free parameters (four in the RAR DM halo plus two of the Freeman disk), adopting the bulge free parameters as detailed in Section \ref{sec:2b}. The predicted circular velocities of the different mass models are denoted with $V(r_i,\mathbf{p})$ and $v_i$ are the observed ones, C is a normalization constant and N is the number of observations. The idea is to fit the free parameters of the model mentioned above to the overall rotation curve (milliparsec inner point $+$ Grand RC).

The implementation of the algorithm was carried out with the help of a tool provided by the PyTorch package \cite{2019arXiv191201703P}, which is an open-source machine learning framework. Such a package was naturally incorporated in the RAR model's code, the last version of which was programmed in PYTHON language (a publicly available version will be published in 2023 as an open code via Github). Since it allows the possibility to include neural networks, it has incorporated the gradient descent method widely used during the training process of such machine learning algorithms. Instead of defining a neural network, we used the backpropagation numerical method, which is able to compute derivatives of compound functions. 
The repeated application of this numerical method has allowed us to apply the gradient descent method to this specific problem.

\begin{figure}[H]
    \includegraphics[width=11.5cm]{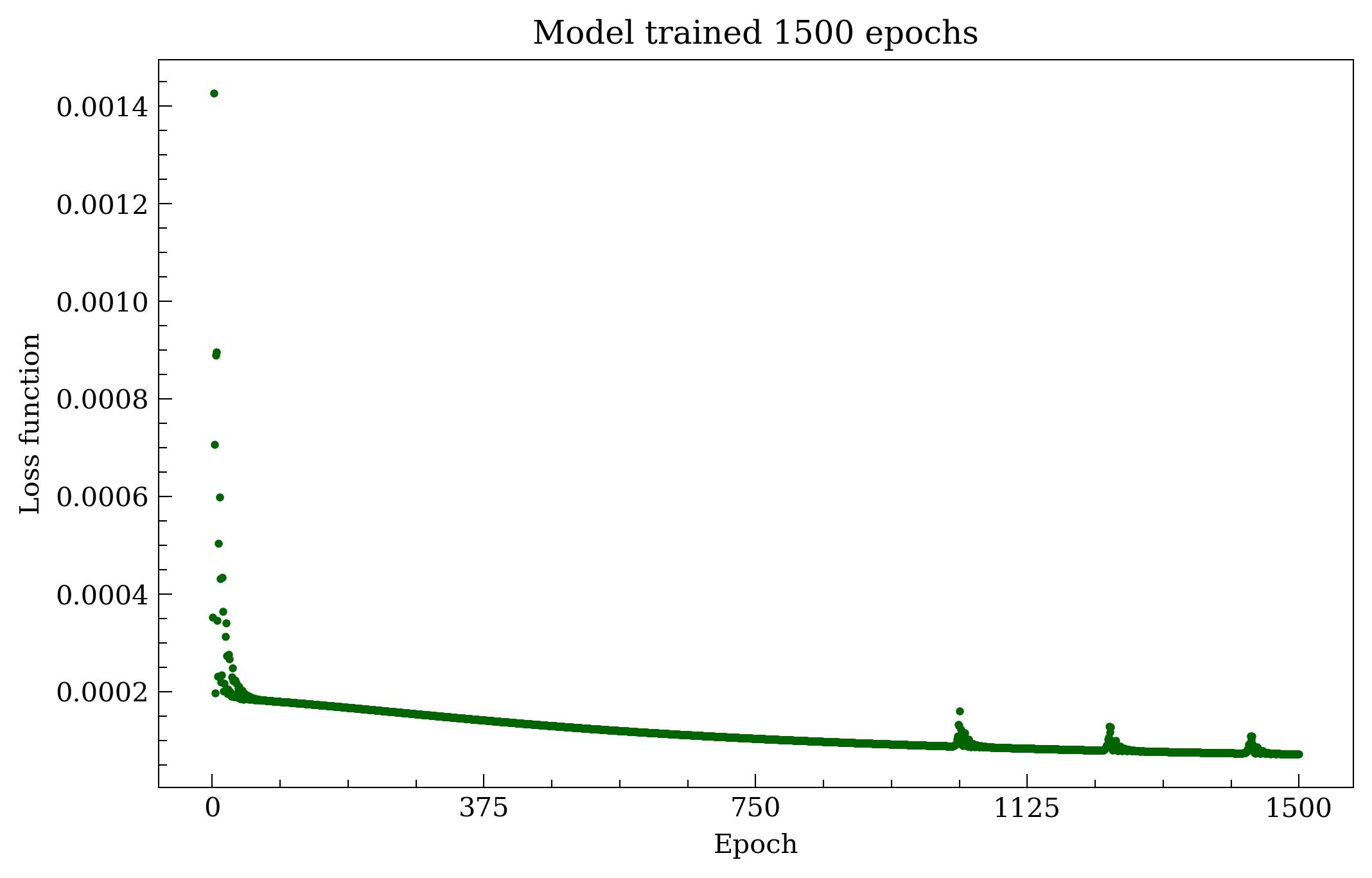}
    \caption{{Loss} 
 function against the step or \textit{epochs} of the method. It can be seen that it is a decreasing function, with little bumps at the tail on the right, indicating a clear minimization trend reaching the value of $0.000071979$ after $1500$ epochs.}
    \label{fig3}
\end{figure}

\section{Results}

\begin{figure}[H]
    \includegraphics[width=10cm]{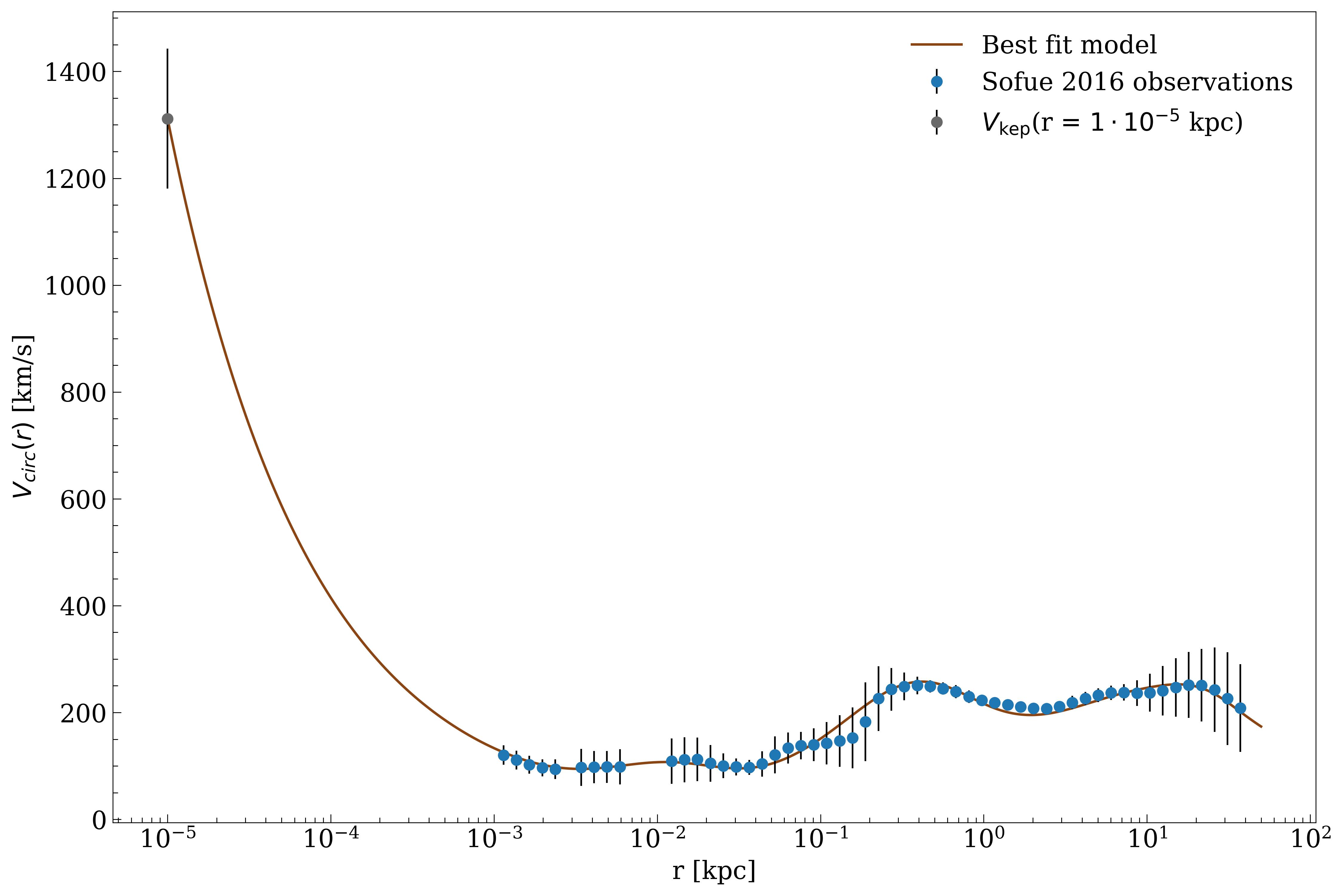}
    \caption{Best-fit circular velocity curve result of the implementation of the gradient descent method. It is remarkable the very good precision achieved in almost all the data-points in few hours CPU time, despite some minor deficiencies at the $\sim 10^{-1}$ kpc scale.}
    \label{fig4}
\end{figure}

In this section, we briefly present the best-fit results to the RC of our Galaxy. The model was iterated with a learning rate of $\gamma=0.001$ and through $1500$ epochs (see Figure \ref{fig3}), giving the following fit (see Figure \ref{fig4}) to the so-called Grand Milky Way RC as given in \cite{2017PASJ...69R...1S}. That is, it covers from inner bulge data points (starting at few pc), to the main bulge data points (peaking at about 0.4 kpc), to data points throughout the disk and DM halo region up to several tens of kpc, all including error bars. In addition, we include independent data points to those analyzed in \cite{2017PASJ...69R...1S}, reaching down to milliparsec scales, and coming from the best resolved S-cluster stars orbiting SgrA* \cite{2017ApJ...837...30G}. 

In the case of Figure \ref{fig4} and for the sake of simplicity, we have included for the S-stars a single data point located at $0.01$ pc, corresponding to the average circular velocity of a typical S-cluster star which falls along the Keplerian velocity trend caused by the supermassive central object. It is important to remark that the behaviour of the circular RC of the Milky Way as predicted by the RAR model (without the central BH) between the innermost data point for a typical S-cluster star at $0.01$ pc and the first data point provided in \cite{2017PASJ...69R...1S}, is Keplerian (i.e., $V_{circ}\propto r^{-1/2}$) all the way to the radius of gravitational influence of the central object at $\sim 1$ pc, as expected (see also \cite{2018PDU....21...82A}) for an analogous plot including for the eight best-resolved S-cluster stars).

It can be seen above that the model generates an excellent fit throughout the whole Galaxy range. The loss value at the last training epoch is $0.000071979$ (see Figure \ref{fig3}). The full best-fit parameters obtained from the iterated model and are given in Table \ref{Table1} below.


\begin{table}[H] 
 \caption{{Best-fit} 
 parameters results after applying the gradient descent method to the set of Milky Way observables under the gravitational potential model described in the main text.}
    \newcolumntype{C}{>{\centering\arraybackslash}X}
    \begin{tabularx}{\textwidth}{CCC}
    \toprule
    \textbf{Parameter}	& \textbf{Seed Value}	& \textbf{Final Value}\\
    \midrule
    m [keV/c$^2$]		& $56.0$			            & $54.809$\\
    $\theta_0$          & $37.766$			        & $37.809$ \\
    $W_0$               & $66.341$                    & $66.449$ \\
    $\beta_0$           & $1.1977\times 10^{-5}$      & $1.1139\times 10^{-5}$\\
    $\Sigma_d$ [$M_\odot$/kpc$^2$] &  $5.9658\times 10^{8}$ &  $1.0882\times 10^{9}$ \\
    $a_d$ [kpc]                &$4.9$                 & $3.0039$ \\
    \bottomrule
    \end{tabularx}
   
\label{Table1}
\end{table}

From the above Table we can see that the final (best-fit) values of the disc parameters are in line with those reported in \cite{2013PASJ...65..118S,2017PASJ...69R...1S}, though the resulting disc here is somewhat more massive with differences of up to $\sim$25\% or $\sim$50\% in the surface DM density parameter respectively. Such differences in the barionic free parameters are expected, since in \cite{2013PASJ...65..118S,2017PASJ...69R...1S} it was assumed to have a NFW profile which is of a cuspy nature (the RAR is cored) and the RAR outer density tail is politropic while the NFW one is a power law (see \cite{2023ApJ...945....1K} for an extensive comparison of possible RAR density morphologies with respect to other typical DM density profiles considered in the literature). 

Regarding the overall DM distribution, the best-fit DM free-parameters imply a compact fermion-core of mass $M_c=3.46\times10^6 M_\odot$ and radius $r_c=4.54\times10^{-4}$ pc (see Figure~\ref{fig1}), thus well within the pericenter of the closest and best-resolved S-cluster star: the S-2 star, which has indeed served as the best (and most accurate) case to constrain the mass of the supermassive dark compact object in SgrA* \cite{2009ApJ...707L.114G}. This is totally in line with the recent results reviewed in \cite{2023Univ....9..197A} within a different phenomenological approach (i.e., fixing only three boundary halo mass conditions from observations as explained in Sec. \ref{sec:1}), where the ability of the RAR model to explain the astrometric data of the S-cluster stars, including the relativistic effects (gravitational redshift and orbit precession) of the S-2 star, was explicitly demonstrated. For the total mass of the Galaxy, we obtain from this work $M_{tot}\approx 3.4\times10^{11} M_\odot$ in line with the results obtained in \cite{2014MNRAS.445.3788G} and in \cite{2018PDU....21...82A} (see footnote 1 in the latter for  further discussion).
Finally, we report a local DM density at the Sun (i.e., at $R_0=8$ kpc) of $0.53$ GeV/cm$^3$, which is well within the $2\sigma$ value as reported in \cite{2013JCAP...07..016N} for another cored (i.e., Burkert) DM profile.





\section{Discussion and Conclusions}

We have implemented a state-of-the-art machine learning numerical technique to best-fit an RC of disc galaxies which include non-analytic DM density profiles. This numerical tool uses a gradient descent method used in the backpropagation process of training a neural network, which can easily include several free-parameters and provide a best-fit within a few hours of CPU-time, thus improving (in some aspects) on other best-fitting methods. As an example study, we have chosen the well-investigated case of our own Galaxy by Y. Sofue \cite{2017PASJ...69R...1S} within the so-called Grand Rotation Curve, and we have further included for milliparsec velocity data coming from the S-cluster star orbiting SgrA* (thus covering a wide range of scales from $10^{-2}$ pc all the way to $10^5$ pc). A key advantage of the machine learning tool used here (i.e., the gradient descent method implemented through Pytorch) is that it can achieve  an excellent accuracy in best-fitting an RC under no-analytic DM halo models, such as the RAR halo model, together with baryonic mass models with varying free parameters within a few hours' time. The relevance of using this kind of DM profile instead of those others commonly used in the literature relies on the fact it is a semi-analytical model built from first-principle physics, such as (quantum) statistical mechanics and thermodynamics, whose more general density profile has a \textit{dense core} -- \textit{diluted halo} which depends on the DM particle mass. As recently shown in \cite{2018PDU....21...82A,2023Univ....9..197A} and references therein, and further verified here, the dark and compact fermion-core can work as an alternative to the central black hole in SgrA* when including data at milliparsec scales from the S-cluster stars.







\authorcontributions{
 Conceptualization, C.R.A. and S.C.; methodology, C.R.A. ans S.C.; software, S.C.; validation, C.R.A.; formal analysis, S.C.; investigation, C.R.A. and S.C.; resources, C.R.A.; data curation, S.C.; writing---original draft preparation, C.R.A.; writing---review and editing, C.R.A. and S.C.; visualization, S.C.; supervision, C.R.A.; project administration, C.R.A.; funding acquisition, C.R.A. All authors have read and agreed to the published version of the manuscript.}

\funding{C.R.A. and S.C. were supported by CONICET of Argentina, and the ANPCyT (grant PICT-2018-03743).}

\dataavailability{The dataset used in this work for the Grand RC was taken from the Y. Sofue's public database \url:{http://www.ioa.s.u-tokyo.ac.jp/~sofue/htdocs/2017paReview/} (accessed on, 15 July 2022) while the innermost central data point at milliparsec scale was taken from the analysis of the S-cluster stars (average circular velocity calculation) as done in \cite{2018PDU....21...82A}.}

\acknowledgments{
C.R.A acknowledges support from ICRANet.}

\conflictsofinterest{The authors declare no conflicts of interest.}




\begin{adjustwidth}{-\extralength}{0cm}
\printendnotes[custom]

\reftitle{References}

\PublishersNote{}
\end{adjustwidth}
\end{document}